\documentstyle[aps,eqsecnum,epsf]{revtex}
\begin{document}
\title{Weak gravity in DGP braneworld model}
\author{Takahiro Tanaka\footnote
{E-mail: tanaka@yukawa.kyoto-u.ac.jp}}
\address{Yukawa Institute for Theoretical Physics, 
Kyoto University, Kyoto 606-8502, Japan}
\maketitle
\begin{abstract}
 We analyze the weak gravity in the braneworld model proposed by 
Dvali-Gabadadze-Porrati, in which the unperturbed background 
spacetime is given by five dimensional Minkowski bulk with 
a brane which has the induced Einstein Hilbert term. 
This model has a critical length scale $r_c$. 
Naively, we expect that the four dimensional general 
relativity (4D GR) is approximately recovered at the scale below $r_c$. 
However, the simple linear perturbation does not work in this regime. 
Only recently the mechanism to recover 4D GR was clarified under the 
restriction to spherically symmetric configurations, and the 
leading correction to 4D GR was derived. 
Here, we develop an alternative formulation which can 
handle more general perturbations. 
We also generalize the model by adding bulk cosmological 
constant and the brane tension. 
\end{abstract}

\section{introduction}

A braneworld model, whose gravity behaves as four dimensional at 
short distance scale but it shows higher dimensional nature 
at larger distances, was proposed by Dvali, Gabadadze, and 
Porrati (DGP)\cite{DGP}. In this model, the brane, on which the fields 
of the standard model are confined, has the induced Einstein-Hilbert 
term\cite{akama,Rubakov:1983bz}. 
This model has various cosmologically interesting 
features\cite{Dvali:2000xg,Dvali:2001gm,Deffayet:2001aw,Deffayet:2001pu,Deffayet:2002sp,Dvali:2001ae,Dvali:2002pe}. 
Particularly in the model with five dimensional bulk, 
an interesting cosmological solution was found, 
in which accelerated expansion of 
the universe at late epoch is realized 
without introducing the cosmological constant\cite{Deffayet:2001pu}. 
Based on this model, a novel mechanism that dilutes 
the cosmological constant was also proposed\cite{Dvali:2002pe}.   

Although we have mentioned above that 
the gravity in this model at short distances 
is expected to behave as four dimensional, 
it is not so transparent if the model actually 
mimics four dimensional general relativity (4D GR).  
The linear analysis of this model shows that the tensor structure 
of the induced metric perturbations takes the five dimensional 
form even at short distance\cite{DGP}. 
The situation is analogous to 
the case of models with massive gravitons.  
In this case the deviation from 4D GR 
does not vanish even in the massless limit, which is 
known as the van Dam-Veltoman-Zakharov discontinuity\cite{vDV,Zakharov}.  
In this context, the possibility that the 4D GR
is recovered by non-linear effect was suggested by\cite{Vainshtein}. 
There have been many discussions about this issue\cite{Damour}. 
We have in particular a clear statement 
that the discontinuity disappears when we 
introduce the cosmological constant\cite{Higuchi,KMP2001}. 
Although the analysis with the cosmological constant is quite suggestive, 
the discontinuity is absent only when the limit is taken keeping 
the length scale determined by the cosmological constant 
much smaller than the compton wavelength of the massive graviton. 
The length scale determined by the cosmological constant 
must be longer than the Hubble horizon size. 
Hence, the recovery of 4D GR cannot be proven by introducing 
negligibly small cosmological constant as far as 
the graviton mass is not completely negligibly small. 

Also specialized to the context of five dimensional DGP model, 
there are various works aiming at answering the question whether 
4D GR is recovered at short distances or not,
and many evidences that indicate the recovery of 4D GR 
were reported\cite{Deffayet:2001uk,Lue:2001gc,Porrati2002,Gruzinov,new}. 
It was shown that the evolution 
equation for homogeneous isotropic universe becomes identical 
to that for 4D GR when the Hubble expansion 
rate is much larger than the inverse of the critical length scale, 
$r_c^{-1}$\cite{Deffayet:2001uk}. 
In Ref.~\cite{Porrati2002}
it was clearly shown that the linear analysis breaks down 
at scale shorter than $(r_c^2 r_g)^{1/3}$ since the brane bending 
becomes non-linear there. Further, a consistent form of a 
black hole metric induced on the brane was presented. 
Approximate black hole solution including the bulk was constructed 
under the restriction to spherically symmetric 
configurations\cite{Gruzinov}. 
The same paper also gave the leading order correction to 4D GR, which 
is potentially observable by the future development of precision 
measurement of solar system\cite{Dvali:2002vf,new}. 
The results in Ref.~\cite{Gruzinov} were extended to the case with 
the background of expanding universe\cite{new}.

In this paper, we develop an alternative formalism which can handle general 
perturbations in weak gravity regime. To handle general perturbations, 
we restrict our consideration to the case that the unperturbed 
metric on the brane is given by Minkowski spacetime. 
We also make further generalization to the model that 
also takes into account the 
bulk cosmological constant and the brane tension balanced 
with it. (Such generalized DGP model was discussed before 
in Refs.~\cite{Collins:2000yb,Shtanov:2000vr,Sahni:2002dx,Park}.) 
We confirm the recovery of 4D GR at short distances and rederive
the leading order correction to it. 

\section{setup}
The model that we consider is defined by the five dimensional action
\begin{equation}
  S={M_4^2\over 4 r_c}\int d^5x\, \sqrt{-g}
      \left(R^{(5)}+{12\over \ell^2}\right) 
    +\int d^4x\, \sqrt{-g^{(4)}}\left({M_4^2\over 2} R^{(4)}
    -{3M_4^2\over r_c\ell}+L_{matter}\right), 
\end{equation}
where $M_4, r_c$ and $\ell$ are constants. 
$R^{(5)}$ and $R^{(4)}$ are, respectively, 
the curvature scalars corresponding to 
the five dimensional metric $g_{\mu\nu}$ and the four dimensional 
one $g^{(4)}_{\mu\nu}$ induced on the brane. 
Here, we added both the bulk cosmological constant and the 
brane tension terms to the original DGP model. They are tuned to admit 
the Minkowski brane as a vacuum solution. 
The model is reduced to the original one 
by setting $\ell\to\infty$. 
The unperturbed background geometry is given by 
five dimensional anti-de Sitter spacetime   
\begin{equation}
 ds^2= g^{(0)}_{ab}dx^a dx^b=dy^2+\gamma_{\mu\nu}(y)dx^\mu dx^\nu
     =dy^2+e^{-2y/\ell}\eta_{\mu\nu}dx^\mu dx^\nu,
\end{equation}
with a brane located at $y=0$, where $Z_2$-symmetry is imposed. 
Here $\eta_{\mu\nu}$ is 4 dimensional 
Minkowski metric. 

\section{semi non-Linear perturbations}
We follow the method of Ref.~\cite{GT} introduced for the purpose 
of analyzing weak gravity in the Randall-Sundrum model~\cite{RS}. 
We prepare two coordinate systems. In the coordinates $\{x^a\}$, 
the gauge is chosen so that the metric perturbations $h_{ab}$ can be 
easily computed in the five dimensional bulk. 
Namely, we use the Randall-Sundrum gauge,
\begin{equation}
  h_{5a}=0,\qquad   h_{\mu}^{~\mu}=0,\qquad 
  h_{\mu~,\nu}^{~\nu}=0. 
\end{equation}
In this paper the fifth direction is the direction of 
extra-dimension. The greek and latin indices 
represent four and five dimensional coordinates, respectively. 
The other coordinate system $\{\bar x^a\}$ satisfies the 
Gaussian normal conditions 
\begin{equation}
  \bar h_{5a}=0,
\end{equation}
and also keeps the location of the brane unperturbed at $\bar y=0$. 
Under the coordinate transformation $x^a=\bar x^a-\xi^a(\bar x)$, 
the metric perturbation transforms as 
\begin{eqnarray}
 \bar h_{ab} & = & h_{ab}(\bar x-\xi(\bar x))
      +\left(
      -g^{(0)}_{ab,5}\xi^5 +{1\over 2} g^{(0)}_{ab,55}(\xi^{5})^2 
          -\cdots\right)\cr
    && -\left\{ \xi^c_{~,a}\left(
      g^{(0)}_{cb}(\bar x-\xi)
         + h_{cb}(\bar x-\xi))\right)+
      \left(a\leftrightarrow b\right)
      \right\} +\left(g^{(0)}_{cd}(\bar x-\xi)
            +h_{cd}(\bar x-\xi)\right)
            \xi^c_{~,a}\xi^d_{~,b}~.
\end{eqnarray}
The argument of the variables is supposed to be $\bar x$ unless 
otherwise is specified, and ``${}_{,a}$'' denotes a differentiation 
with respect to $\bar x^a$. 

The conditions that 
the $\{00\}$ component and $\{0\mu\}$ components are 
zero in both coordinates provide 
equations for the gauge parameters, which are solved up to 
second order as  
\begin{eqnarray}
  \xi^5 & = & \hat \xi^5 + \stackrel{(2)}\xi{}\!\!^5,
\cr
  \xi^\mu & = & {\ell\over 2}(\eta^{\mu\nu}-\gamma^{\mu\nu}) 
           \hat \xi^5_{~,\nu} 
           +\hat\xi^\mu + \stackrel{(2)}\xi{}\!\!^\mu,
\end{eqnarray}
where $\hat \xi^5(\bar x^\rho)$ and $\hat \xi^\mu(\bar x^\rho)$ 
are the values of the 
gauge parameters evaluated on the brane, and 
\begin{eqnarray}
 \stackrel{(2)}\xi{}\!\!^5 & = & \int_0^{\bar y} d\bar y
          \, \gamma^{\mu\nu}\hat\xi^5_{~,\mu}\hat\xi^5_{~,\nu},\cr
 \stackrel{(2)}\xi{}\!\!^\mu & = & \int_0^{\bar y} d\bar y
          \, \gamma^{\mu\rho}\left[\hat\xi^5_{~,\nu}
             \left(\bar h_\rho^{~\nu}+{2\over \ell} 
                \delta_\rho^{~\nu}\hat\xi^5\right) 
             -\xi^\sigma_{~,\rho} 
             \hat\xi^5_{~,\sigma}-\stackrel{(2)}\xi{}\!^5_{~,\rho}
                 \right].
\end{eqnarray}
We assume the following order counting,
\begin{equation}
  \hat\xi^5_{,\mu}\lesssim \epsilon,
\quad
  {\hat\xi^5\over \ell},   {\hat\xi^5\over r_c}, 
  \hat\xi^\rho_{~,\mu}, \left. \bar h^\rho_{~\mu}\right\vert_{\bar y=0}
  \lesssim \epsilon^2,
\end{equation}
and keep the terms up to $O(\epsilon^2)$. 
Here $\epsilon^2$ is the order of the Newton potential 
$\Phi=-{1\over 2}\bar h_{00}$.
Later we will 
verify the consistency of this order counting. 
Then, the transformation for $\{\mu\nu\}$ components reduces to 
\begin{equation}
 \bar h_{\mu\nu}(\bar x)=h_{\mu\nu}(\bar x-\xi(\bar x))+ \delta h_{\mu\nu},
\label{transfin}
\end{equation}
with 
\begin{eqnarray}
 \delta h_{\mu\nu} =
      {2\over \ell}\gamma_{\mu\nu}\xi^5 
        - \xi_{\mu,\nu} - \xi_{\nu,\mu}
        + \hat \xi^5_{~,\mu}\hat \xi^5_{~,\nu}. 
\end{eqnarray}
Hereafter, the Greek indices are lowered or raised by 
the metric $\gamma_{\mu\nu}$.  

The brane location is given by $\bar y=0$. 
Hence in the $\{x^a\}$-coordinates 
the brane is bent. 
For simplicity, we impose the harmonic gauge condition, 
$\bar h_{\mu~,\nu}^{~\nu}={1\over 2}\bar h_{,\mu}$ for the 
induced metric on the brane.   
To the second order in $\epsilon$, this condition gives 
\begin{equation}
 \hat\xi_\mu=\Box^{-1}\left[
    -{2\over \ell}\hat \xi^5_{,\mu}
    +\left(\Box\hat\xi^5\right)\hat\xi^5_{,\mu}\right]. 
\label{hatximu}
\end{equation}
From this relation, we find that the assumption 
$\hat \xi^\mu_{~,\nu}=O(\epsilon^2)$ is 
consistent if the assumed order of $\hat\xi^5$ is correct. 
Substituting~(\ref{hatximu}), 
the gauge transformation $\delta h_{\mu\nu}$ 
evaluated on the brane becomes 
\begin{equation}
 \delta h_{\mu\nu}\vert_{\bar y=0}={2\over \ell}\gamma_{\mu\nu}\hat\xi{}^5
   +\Box^{-1}\left[{4\over \ell}\hat \xi{}^5_{,\mu\nu}
    +2\gamma^{\rho\sigma}\hat\xi{}^5_{,\rho\mu}\hat\xi{}^5_{,\sigma\nu}
    -2(\Box\hat\xi{}^5)\hat\xi{}^5_{,\mu\nu}\right].
\end{equation}
Then, the trace of the induced metric is also evaluated as 
\begin{eqnarray}
 \bar h 
  & = & {12\over \ell}\hat\xi^5+2\Box^{-1}
   \left(\gamma^{\mu\nu}\gamma^{\rho\sigma}
       \hat\xi{}^5_{,\mu\rho} \hat\xi{}^5_{,\nu\sigma}
        -(\Box\hat\xi{}^5)^2\right). 
\end{eqnarray}

Next we consider the junction condition. 
After a straightforward calculation, we can show
\begin{equation}
 \left(\partial_{\bar y}+2\ell^{-1}\right)\bar h_{\mu\nu}
    = \left(\partial_{\bar y}+2\ell^{-1}\right)h_{\mu\nu}(\bar x-\xi)
      +2\hat\xi^5_{,\mu\nu}+{2\over \ell}\Xi_{\mu\nu}~,
\qquad ({\rm at}~~\bar y=0),
\end{equation}
with
\begin{equation}
 \Xi_{\mu\nu}:= \gamma_{\mu\nu}
          \gamma^{\rho\sigma}\hat\xi^5_{,\rho}\hat\xi^5_{,\sigma}
          +\hat\xi^5_{,\mu}\hat\xi^5_{,\nu}~. 
\end{equation}
Using this relation, the junction condition becomes 
\begin{equation}
T_{\mu\nu}-{1\over 2}\gamma_{\mu\nu} T 
  +{M_4^2\over 2} \Box \bar h_{\mu\nu}
=-{M_4^2\over 2r_c}\left[
  \left.\left(\partial_{\bar y}+2\ell^{-1}\right)h_{\mu\nu}(\bar x-\xi)
  \right\vert_{\bar y=0}+2\hat\xi^5_{,\mu\nu}+{2\Xi_{\mu\nu}\over \ell}+
      \gamma_{\mu\nu}\left(\Box\hat\xi^5 + {\Xi\over \ell}\right)\right].  
\label{eq:junction}
\end{equation}

The equation that determines the brane bending is obtained from 
the trace of the above equation as 
\begin{eqnarray}
 \Box^{-1}{1\over M_4^2} T & = & {\bar h\over 2} +{3\over r_c}\left( 
    \hat\xi^5 +\Box^{-1}{\Xi\over \ell}\right)\cr
  & = & {3\over r_c^*} \hat\xi^5
   +\Box^{-1}\left(\gamma^{\mu\nu}\gamma^{\rho\sigma}
       \hat\xi{}^5_{,\mu\rho} \hat\xi{}^5_{,\nu\sigma}
        -(\Box\hat\xi{}^5)^2\right)+{3\over r_c \ell}\Box^{-1}\Xi,  
\label{bending}
\end{eqnarray} 
where we have introduced $r_c^*:= [({1/ r_c})+({2/ \ell})]^{-1}$. 
We can neglect the last term on the right hand side of 
Eq.~(\ref{bending}) because it is always higher order 
compared with the first term. The left hand side is something like  
the Newton potential $\Phi$, hence we assume it to be $O(\epsilon^2)$. 
Outside the matter distribution with the total mass $m$, 
the left hand side can be expressed as $\approx r_g/r$, where 
$r_g:=m/4\pi M_4^2$. 
At large scale, $r\gtrsim (r_g r_c^2)^{1/3}$, 
the first term on the right hand side dominates, while at small 
scale, $r\lesssim (r_g r_c^2)^{1/3}$, the second term dominates. 
Therefore we have
\begin{equation} 
O(\epsilon^2)=\max \left(
    \left\vert{\hat\xi^5\over r_c^*}\right\vert, 
    \left\vert\hat\xi^5_{~,\mu}\right\vert^2 \right). 
\end{equation}
Thus we find that our assumption as to the order counting for
$\hat\xi^5$ is justified. 

\section{mechanism for recovering four dimensional general relativity}

The remaining task is to evaluate 
$\left(\partial_{\bar y}+2\ell^{-1}\right) h_{\mu\nu}(\bar x-\xi(\bar x))
  \vert_{\bar y=0}$. 
Here we need to solve the bulk field equations. Different from the 
R-S case, we solve the bulk equations with the Dirichlet 
boundary condition (\ref{transfin}). Here, we note that 
the location of the brane is not a straight sheet in the 
coordinates in the R-S gauge $\{x^a\}$. 

We can give the general solution for 
the bulk field equations as a superposition of homogeneous mode
solutions with purely outgoing boundary condition: 
\begin{eqnarray}
 h_{\mu\nu}(x)
    & = & \int {\cal H}_{\mu\nu}(p) e^{ip_{\mu}x^\mu} 
           K_2(p\ell e^{y/\ell})\, d^4p\cr
    & = & \int {\cal H}_{\mu\nu}(p) e^{ip_{\mu}(\bar x^\mu-\xi^\mu(\bar x))}
              K_2(p\ell e^{\bar y-\xi^5(\bar x)/\ell})\, d^4p~.
\end{eqnarray}
where $K_2(p\ell)$ is the modified Bessel function 
and ${\cal H}_{\mu\nu}$ is the expansion coefficient. 
The coefficient ${\cal H}_{\mu\nu}$ is to be determined 
so as to satisfy the Dirichlet boundary condition 
\begin{equation}
 h_{\mu\nu}|_{\bar y=0}(\bar x)=
     \int {\cal H}_{\mu\nu}(p) e^{ip_{\mu}(\bar x^\mu-\xi^\mu(\bar x))}
              K_2(p\ell e^{-\hat \xi^5(\bar x)/\ell})\, d^4p~.
\end{equation}
If we are allowed to approximate the above expression by setting
$\xi^a=0$, we have $\tilde h_{\mu\nu}(p):=
  (2\pi)^{-4}\int d^4\bar x\, e^{-ip_\rho \bar x^\rho} h_{\mu\nu}(\bar
x) = {\cal H}_{\mu\nu}(p)$, and therefore we have  
\begin{eqnarray}
(2\pi)^{-4}\int d^4\bar x\, e^{-ip_\rho \bar x^\rho}
  \left.\left(\partial_{\bar y}+2\ell^{-1}\right) 
  h_{\mu\nu}\right\vert_{\bar y=0}(\bar x) = 
  -{pK_1(p\ell)\over K_2(p\ell)}\tilde h_{\mu\nu}(p)
  =-{pK_1(p\ell)\over K_2(p\ell)}
    (\tilde{\bar h}_{\mu\nu}-\delta\tilde h_{\mu\nu}).
\label{crudeapp}
\end{eqnarray}
We think that the errors caused by this naive approximation 
are not large, although any rigorous proof is not ready yet. 
If the leading errors are simply proportional to 
$h_{\mu\nu}\xi$, we can neglect them since they are 
higher order in $\epsilon$. Such a naive expansion with respect to $\xi$ 
will be justified for small $p$. But for large $p$, we will not be 
allowed to expand the combination $p\xi$ in the exponent. 
However, as we will see below, even the leading correction to 
the gravitational potential coming from the contribution of this part 
is suppressed to be irrelevantly small 
at small scale $r\lesssim r_c$. 
Hence, the errors due to this naive approximation can be 
crucial only if this approximation 
significantly underestimate the magnitude of  
$(\partial_{\bar y}+2\ell^{-1}) h_{\mu\nu}|_{\bar y=0}$, 
which is quite unlikely. 

Using~(\ref{crudeapp}), the junction condition (\ref{eq:junction}) 
is written down explicitly as 
\begin{eqnarray}
\tilde{\bar h}_{\mu\nu} -
   {2\over M_4^2}{\cal D}
      \left(\tilde T_{\mu\nu}-{1\over 2}\gamma_{\mu\nu}
      \tilde T\right)
    & = &  {\cal D} \Biggl[
    \left(-{p^2\over  r_c}+{2 p K_1(p\ell)\over r_c \ell
      K_2(p\ell)}\right)
     \gamma_{\mu\nu}\tilde{\hat \xi}{}^5 
    -{2\over r_c} p_\mu p_\nu \tilde{\hat\xi}{}^5 \cr
     & & \qquad
      +{p K_1(p\ell)\over r_c K_2(p\ell)}\delta\tilde h^{(2)}_{\mu\nu}
      +{1\over r_c\ell}\left(\tilde \Xi_{\mu\nu}+{1\over
      2}\gamma_{\mu\nu}\tilde \Xi\right) \Biggr], 
\qquad ({\rm at}\quad \bar y=0),
\label{barhfin}
\end{eqnarray}
where
\begin{equation}
 {\cal D}={1\over p^2+ 
   \displaystyle{p K_1(p\ell)\over r_c K_2(p\ell)}}~, 
\end{equation}
and
\begin{equation}
 \delta h_{\mu\nu}^{(2)}=
     \Box^{-1}\left[
    2\gamma^{\rho\sigma}\hat\xi{}^5_{,\rho\mu}\hat\xi{}^5_{,\sigma\nu}
    -2(\Box\hat\xi{}^5)\hat\xi{}^5_{,\mu\nu}\right]~.
\end{equation}
Here the quantitiy with $\tilde{~}$ represents the Fourier coefficient 
of the corresponding variable as before.
We show that in the square brackets on the right hand side of 
Eq.~(\ref{barhfin}) the first term gives the 
dominant contribution. 
We can drop the last two terms 
simply because they are always higher order in $\epsilon$ 
compared with the first term.  
The second term is irrelevant since it 
can be eliminated by a four-dimensional gauge transformation. 
As a result, the equation that determines the metric induced 
on the brane $\bar h_{\mu\nu}$ is 
reduced to the one for the linear theory. The only difference 
is in the equation that determines the brane bending (\ref{bending}). 

The order of magnitude of the first term on the right hand side 
of Eq.~(\ref{bending}) is estimated as 
$\hat\xi^5/r_c\sim \Phi^{1/2} r/r_c$ at small scale, 
and hence it is suppressed by the factor $\Phi^{-1/2} r/r_c$ 
compared with the Newton potential $\Phi$. 
The leading behavior of the induced metric 
is therefore determined by setting the left hand side of Eq.~(\ref{barhfin}) 
to zero. Thus we conclude that 4D GR is recovered 
by taking into account the non-linear brane bending 
for weak gravity at small scale $r\lesssim (r_c^2 r_g)^{1/3}$.  
If we take the limit $r_c\to\infty$, all length scales come 
into this regime. Hence, we have confirmed 
the absence of van Dam-Veltman-Zakharov 
discontinuity. However, 
because of the factor $\Phi^{-1/2}$ the 
leading order deviation from 4D GR at small scale  
is less strongly suppressed than the naively expected suppression 
$r/r_c$, as first pointed out in Ref.~\cite{Gruzinov}.  

At large scale, this term becomes more and more important. 
For $r\gtrsim (r_c^2 r_g)^{1/3}$, we have 
\begin{equation}
 \tilde {\hat \xi}{}^5\approx -{r_c^* \tilde T\over 3 M_4^2 p^2}.  
\label{linearxi}
\end{equation}
This is nothing but the result for the linearized case. 
In the next section, we discuss the regime 
where the linear theory is valid. 
After that, in the succeeding section, we discuss the leading 
order correction to the 4D GR at short distance scale assuming 
static and spherically symmetric configurations.

\section{linear regime}
In this section we consider 
perturbations at large scale  
$r\gg (r_c^2 r_g)^{1/3}$, where the linear theory is valid. 
Substituting Eq.~(\ref{linearxi}) into Eq.~(\ref{barhfin}), we obtain
\begin{equation}
 \tilde{\bar h}_{\mu\nu} \approx 
   {2\over M_4^2}{\cal D}
      \left(\tilde T_{\mu\nu}-{1\over 2}\alpha \gamma_{\mu\nu}
      \tilde T\right),  
\end{equation}
with 
\begin{eqnarray}
\alpha & = & {r_c +{\ell \over 3}\left(
             1+{1\over \ell p}{K_1(p\ell)\over K_2(p\ell)}\right)
         \over r_c+{\ell\over 2}}. 
\end{eqnarray}

First we look at the behavior of the propagator $D$, which is 
already discussed in Ref.~\cite{Park}. 
When we consider the length scale much smaller than $\ell$,
we have 
\begin{equation}
  {\cal D}\to {1\over p^2+ r_c^{-1}p},\qquad  (p\ell\gg 1). 
\end{equation}
At the length scale smaller than $r_c$ ($p r_c\gg 1$), 
the propagator ${\cal D}$ behaves as that for a four dimensional field. 
On the other hand, at the intermediate scale between 
$r_c$ and $\ell$ ($r_c \gg p^{-1} \gg \ell$), the propagator 
behaves as that for a five dimensional field. 
When the length scale is much larger than $\ell$, 
$K_1(p\ell)/K_2(p\ell)$ goes to $p\ell/2$. 
Thus we have 
\begin{equation}
  {\cal D}\to {2r_c \over (\ell+2r_c) p^2}, \qquad (p\ell\ll 1).  
\end{equation}
Hence again the propagator ${\cal D}$ 
behaves as a four dimensional field, but the Newton's constant is 
not given by $2 M_4^{-2}$ but by $2 M_4^{-2}/(1+\ell/2r_c)$.

Next we turn to the tensor structure specified by 
$\alpha$. For four dimensional 
massless graviton we have $\alpha=1$, while  
$\alpha={2\over 3}$ for the case of massive graviton.   
For $p\ell \gg 1$, we have 
\begin{equation}
 \alpha \to {1+{\ell\over 3r_c}\over 1+{\ell\over 2r_c}}~.
\end{equation}
We have $\alpha\to 1$ for $\ell \ll r_c$, 
while $\alpha\to 2/3$ for $\ell \ll r_c$. 
On the other hand, for $p\ell \ll 1$, we have $\alpha\to 1$ irrespective of 
the ratio between $\ell$ and $r_c$. 

The results are summarized in Fig.1. When $r\gtrsim \ell$, the 4D GR 
is realized by the Randall-Sundrum mechanism. The effective 
Planck mass differs from $M_4$ in this case. On the other hand, 
when $r\lesssim r_c$, the gravity becomes four dimensional again, but 
the tensor structure differs from 4D GR.  

\begin{figure}
\centerline{
\epsfysize4cm\epsfbox{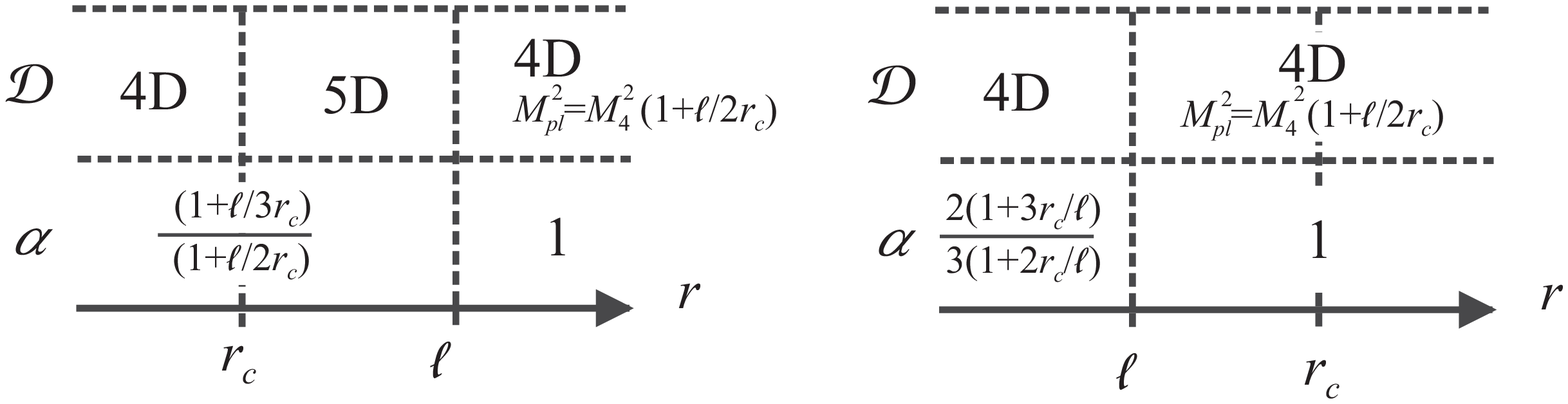}
}
\caption{Summary of the results of linear analysis, which is 
valid when $r\gg (r_c^2 r_g)^{1/3}$.  
The horizontal axis represents length scale. The raws labeled 
${\cal D}$ and $\alpha$, respectively, 
explain the property of the propagator ${\cal D}$  
and the indicator of the tensor structure $\alpha$. 
}
\end{figure}
\section{static spherical symmetric non-relativistic star}
\label{sec:static}
We consider a static spherical symmetric non-relativistic star. 
Assuming that 
the energy momentum tensor is dominated by the $\{00\}$-component 
$T_{00}=\rho$, we neglect the effect of pressure when we solve 
the metric perturbation. 
First we solve non-linear equation  for $\hat\xi^5$ (\ref{bending}).
Under the present assumptions, Eq.~(\ref{bending}) is simplified as 
\begin{equation}
 -{1\over M_4^2} \rho ={1\over r^2}\partial_r 
              \left({3r^2\over r_c^*}\hat\xi^5_{,r}
                    - 2r(\hat\xi^5_{,r})^2\right).
\end{equation}
This equation can be immediately integrated once, and we obtain 
\begin{equation}
  2(\hat\xi^5_{,r})^2-{3 r\over r_c^*}\hat\xi^5_{,r}-{r_g(r)\over r}=0,
\end{equation}
where
\begin{equation}
 r_g(r)={1\over M_4^2}\int_0^r dr\, r^2\rho. 
\end{equation}
Outside the star, we have $r_g(r)=r_g=m/4\pi M_4^2$. 
Solving the above equation with respect to $\hat\xi^5_{,r}$, we have 
\begin{equation} 
 \hat\xi^5=\int dr \, 
      \left({3r\over 4r_c^*}-{1\over 4}\sqrt{\left({3r\over r_c^*}\right)^2+
       {8 r_g(r)\over r}}\right). 
\end{equation}
Here we have chosen the signature in front of the square root 
 imposing the condition that 
$\hat\xi^5_{,r}$ does not become large at $r\to\infty$. 
The other branch with ``+'' sign is outside the scope of the present 
formalism since we have assumed the Minkowski brane background. 
At small scale, this expression reduces to 
$\displaystyle
 \hat\xi^5\approx -\int dr \, 
    \sqrt{r_g(r)/ 2r}. 
$
Outside the matter distribution, we simply have 
$\hat\xi^5=-\sqrt{2r\, r_g}$. 
Hence, the correction to the Newton potential 
is given by 
\begin{equation}
 \delta\Phi \approx \sqrt{{r r_g\over 2r_c^2}}~,   
\end{equation}
which recovers the result obtained in Ref.~\cite{Gruzinov}.

\section{conclusion}
We developed a formalism to calculate 
the metric perturbations induced by the matter localized 
on the brane 
in the generalized DGP model, in which the bulk cosmological 
constant and the brane tension terms are added. 
Here we clarified the mechanism for disappearing the 
van Dam-Veltman-Zakharov 
discontinuity in this model. 
In our approach, the crucial point was to take into 
account a part of second order perturbations of the brane bending. 
This method was largely motivated by the recent works by 
\cite{Porrati2003,Rubakov2003}.
Our scheme almost completely controls the order of magnitude of 
the neglected higher order correction terms. 
In this sense, we think that this work gives an alternative 
sufficiently satisfactory 
proof of the absence of the van Dam-Veltman-Zakharov 
discontinuity in the DGP model. 
Under the restriction to the static and spherically symmetric source,  
we confirmed that our formulation correctly reproduces the 
leading order correction to 4D GR at short distances obtained 
in \cite{Gruzinov}. 

\section*{Acknowledgments} 
The author would like to thank T.~Wiseman for useful suggestions 
and discussions. To complete this work, the discussion during and 
after the YITP workshop YITP-W-02-19 was useful. 
This work was supported in part by the Monbukagakusho Grant-in-Aid 
No.~14740165.


\begin{thebibliography}{ederf}

\bibitem{DGP}
G.~R.~Dvali, G.~Gabadadze and M.~Porrati,
Phys.\ Lett.\ B {\bf 485}, 208 (2000).

\bibitem{akama}
K.~Akama, Lect. Notes Phys. {\bf 176}, 267 (1982). 

\bibitem{Rubakov:1983bz}
V.~A.~Rubakov and M.~E.~Shaposhnikov,
Phys.\ Lett.\ B {\bf 125}, 139 (1983).


\bibitem{Dvali:2000xg}
G.~R.~Dvali and G.~Gabadadze,
Phys.\ Rev.\ D {\bf 63}, 065007 (2001).


\bibitem{Dvali:2001gm}
G.~R.~Dvali, G.~Gabadadze, M.~Kolanovic and F.~Nitti,
Phys.\ Rev.\ D {\bf 64}, 084004 (2001).

\bibitem{Deffayet:2001aw}
C.~Deffayet, G.~R.~Dvali, G.~Gabadadze and A.~Lue,
Phys.\ Rev.\ D {\bf 64}, 104002 (2001).

\bibitem{Deffayet:2001pu}
C.~Deffayet, G.~R.~Dvali and G.~Gabadadze,
Phys.\ Rev.\ D {\bf 65}, 044023 (2002).

\bibitem{Deffayet:2002sp}
C.~Deffayet, S.~J.~Landau, J.~Raux, M.~Zaldarriaga and P.~Astier,
Phys.\ Rev.\ D {\bf 66}, 024019 (2002).

\bibitem{Dvali:2001ae}
G.~Dvali, G.~Gabadadze, X.~r.~Hou and E.~Sefusatti,
Phys.\ Rev.\ D {\bf 67}, 044019 (2003).

\bibitem{Dvali:2002pe}
G.~Dvali, G.~Gabadadze and M.~Shifman,
Phys.\ Rev.\ D {\bf 67}, 044020 (2003).

\bibitem{vDV}
H. van Dam and M.~Veltman,
Nucl. Phys. {\bf B22}, 397 (1970).

\bibitem{Zakharov}
V. I. Zakharov, JETP Lett. {\bf 12}, 312 (1970).

\bibitem{Vainshtein}
A.I.~Vainshtein, Phys. Lett. {\bf 39B}, 393 (1972). 

\bibitem{Damour}
{\it e.g.} T.~Damour, I.~I.~Kogan and A.~Papazoglou,
Phys.\ Rev.\ D {\bf 67}, 064009 (2003).

\bibitem{Higuchi}
A. Higuchi, Nucl.\ Phys.\ {\bf B282}, 397 (1987); ibid. 
{\bf B325}, 745 (1989).

\bibitem{KMP2001}
I.I.~Kogan, S.~Mouslopoulos, and A.~Papazoglou, 
Phys.\ Lett.\ {\bf B503}, 173 (2001).

\bibitem{Deffayet:2001uk}
C.~Deffayet, G.~R.~Dvali, G.~Gabadadze and A.~I.~Vainshtein,
Phys.\ Rev.\ D {\bf 65}, 044026 (2002).

\bibitem{Lue:2001gc}
A.~Lue,
Phys.\ Rev.\ D {\bf 66}, 043509 (2002).

\bibitem{Porrati2002}
M.~Porrati, 
Phys.\ Lett.\ {\bf B534}, 209-215 (2002). 

\bibitem{Gruzinov}
A.~Gruzinov,
astro-ph/0112246. 

\bibitem{new}
A.~Lue and G.~Starkman,
Phys.\ Rev.\ D {\bf 67}, 064002 (2003).

\bibitem{Dvali:2002vf}
G.~Dvali, A.~Gruzinov and M.~Zaldarriaga,
hep-ph/0212069.

\bibitem{Collins:2000yb}
tH.~Collins and B.~Holdom,
Phys.\ Rev.\ D {\bf 62}, 105009 (2000).

\bibitem{Shtanov:2000vr}
Y.~V.~Shtanov,
arXiv:hep-th/0005193.

\bibitem{Sahni:2002dx}
V.~Sahni and Y.~Shtanov,
astro-ph/0202346.

\bibitem{Park}
D.~K.~Park,
arXiv:hep-th/0304056.

\bibitem{GT}
J.~Garriga and T.~Tanaka, 
Phys. Rev. Lett.{\bf 84}, 2778 (2000).  

\bibitem{RS}
 L.~Randall and R.~Sundrum, Phys. Rev. Lett.{\bf 83}, 
3370 (1999); ibid. {\bf 83}, 4690 (1999).





\bibitem{Porrati2003}
M.~A.~Luty, M.~Porrati and R.~Rattazzi,
hep-th/0303116.

\bibitem{Rubakov2003}
V.~A.~Rubakov,
hep-th/0303125.
\end{thebibliography}
\end{document}